\documentclass[aps,prb,twocolumn,showpacs]{revtex4}
\usepackage{amsmath}
\usepackage{graphicx}
\usepackage{units}  
\usepackage{xspace}
\usepackage{subfigure}
\usepackage{hyperref}
\newcommand{\beq}{\begin{equation}}
\newcommand{\eeq}{\end{equation}}
\newcommand{\bea}{\begin{eqnarray}}
\newcommand{\eea}{\end{eqnarray}}

\newcommand{\sx}{\sigma_{ x}}
\newcommand{\sy}{\sigma_{ y}}
\newcommand{\sz}{\sigma_{ z}}
\newcommand{\LLo}{\ensuremath{\textrm{LL}_0}\xspace}
\newcommand{\moire}{moir\'{e}\xspace}
\newcommand{\Moire}{Moir\'{e}\xspace}
\newcommand{\vf}{v}

\usepackage[usenames]{color}

\begin{document}
\bibliographystyle{apsrev}
 
\title{ Effective theory of rotationally faulted multilayer graphene - the local limit}
\author{M. Kindermann}
\author{P. N. First}
\affiliation{ School of Physics, Georgia Institute of Technology, Atlanta, Georgia 30332, USA }

\date{\today}
\begin{abstract} 
Interlayer coupling in rotationally faulted graphene multilayers breaks the local sublattice-symmetry of the individual layers. Earlier we have presented a  theory of this mechanism, which  reduces to an effective Dirac model with space-dependent mass  in an important limit. It thus makes a wealth of existing knowledge available for the study of rotationally faulted graphene multilayers. Agreement of this theory with a recent experiment in a strong magnetic field was demonstrated. Here  we explore some of the predictions of this theory for the system in zero magnetic field at large interlayer bias, when it becomes local in space. We  use that theory to illuminate the physics of localization and velocity renormalization in twisted graphene bilayers. \end{abstract}

\maketitle 

\section{Introduction}
Experiments indicate that the 10--100 individual graphene layers grown on the carbon-terminated face of SiC are surprisingly well decoupled from one another electronically. Early spectroscopic measurements \cite{sadowski:prl06,orlita:prl08}  found a linear low-energy electronic dispersion to the experimental precision, like that of single-layer graphene \cite{novoselov:nat05,zhang:nat05}. In scanning tunneling microscopy/spectroscopy (STM/STS) measurements the Landau level quantization of the material in a magnetic field was found to be essentially that of single-layer graphene %and distinctly different from conventional 2D electron gases with a quadratic dispersion relation 
 \cite{miller:sci09}. Theoretically it has been shown that this approximate decoupling of different layers is due to a relative twist of the layers with respect to each other \cite{latil:prb07,lopes:prl07,hass:prl08,shallcross:prl08,shallcross:prb10,laissardiere:nal10,mele:prb10}. % Heuristically, the crystal growth process favors twist angles $\theta$ within near $30^\circ$ and within $\approx 4^\circ$ of layer alignment \cite{hass:prl08}.% For the  twist angles that we address in what follows, the interlayer coupling at the dominant wavevectors is perturbative \cite{lopes:prl07}. For those typical rotation angles 
 A renormalization of the electron velocity \cite{lopes:prl07,laissardiere:nal10}, van Hove singularities \cite{andrei:nap10}, and interlayer transport \cite{bistritzer:prb10} have been discussed as residual effects of the interlayer coupling. 
 
In a recent STM measurement on multilayer epitaxial graphene \cite{miller:nap10} a spatially modulated splitting $\Delta\lesssim \unit[10]{meV}$ of the zeroth Landau level (\LLo) was observed. In view of the above this finding is intriguing, since the states forming \LLo of an isolated layer of graphene without electron-electron interactions are degenerate. % and those interactions are expected to be screened at by the highly doped graphene layers close to the SiC substrate. 
Many aspects of the experimental data indicate that this splitting is due to the  coupling between graphene layers.  In Ref.~\onlinecite{miller:nap10} we   proposed a phenomenological theory of the interlayer interaction. In that theory  a   ``staggered'' electric potential (a potential with opposite sign on the two sublattices) breaks the sublattice-symmetry locally. This model qualitatively accounts for the main features of the experimental data.  In Ref.\ \onlinecite{kindermann:prb11}  we have presented a microscopic theory of the interlayer coupling in rotationally faulted graphene multilayers that reduces to the phenomenological model of Ref.~\onlinecite{miller:nap10}  in certain limits. The theory is formulated  for a single layer of graphene and it accounts for the coupling to other layers by effective potentials and an effective mass that are possibly non-local in space. The theory of Ref.\ \onlinecite{kindermann:prb11} accounts for the main  features of the    experimental findings \cite{miller:nap10}, both qualitatively and quantitatively. 

 A   number of  intriguing results have been obtained theoretically in electronic structure calculations of rotationally-faulted multilayer graphene also in zero magnetic field \cite{laissardiere:nal10,bistritzer:pna11}. One may therefore  ask  whether the theory of Ref.\ \onlinecite{kindermann:prb11} can provide an intuitive understanding also of these results, as it did for the physics of the material in high magnetic field: is that theory an advantageous starting point to exploring the physics of rotationally-faulted multilayer graphene also in zero magnetic field? 
 
 In this article we give a partial answer to that question by exploring  predictions of the theory  of Ref.\ \onlinecite{kindermann:prb11} in zero magnetic field for quantities that have displayed interesting features in the calculations  of Ref.\  \onlinecite{laissardiere:nal10} and by seeking an interpretation of the results in qualitative terms.  We focus on the spatially local limit of  the theory  \onlinecite{kindermann:prb11} that corresponds to the phenomenological model of  Ref.~\onlinecite{miller:nap10}: a single-layer Dirac model with oscillating effective potentials and a space-dependent mass. That limit  is realized in the presence of a large interlayer bias.   The theory predicts a density of states in qualitative agreement with  experimental  topographic STM measurements. Moreover, our calculation qualitatively reproduces some of the main observations of the mentioned electronic structure calculations of twisted graphene bilayers \cite{laissardiere:nal10} such as a localization of electronic states and a corresponding velocity suppression. The agreement is not quantitative, since  the  calculations of Ref.\  \onlinecite{laissardiere:nal10}  were not obtained in the spatially local limit assumed here. But  in the framework of the theory of Ref.\ \onlinecite{kindermann:prb11}  these predictions do have an intuitive explanation in terms of known results about the Dirac equation with a space-dependent mass.
This suggests that this theory is indeed an advantageous starting point for the exploration of the physics of rotationally faulted graphene multilayers also in zero magnetic field.

We start our discussion with Section \ref{mod}, where we restate the model on which our earlier theory \cite{kindermann:prb11} is based. In Section \ref{loc} we take the limit of a large interlayer bias, when the effective theory of Ref.\ \onlinecite{kindermann:prb11} becomes local in space. We then proceed to evaluate the   density of states and the electron velocity renormalization predicted by this theory in zero magnetic field. We do that first perturbatively in the interlayer coupling in Section \ref{pert}. In Section \ref{nonpert} we  then analyze nonperturbatively a toy model that resembles the original theory, but assumes a  simplified spatial structure of the effective staggered potential. We conclude in Section \ref{conclusion}.
 
 \section{Model} \label{mod}
 
\begin{figure}[h]
\includegraphics[width=0.85\linewidth]{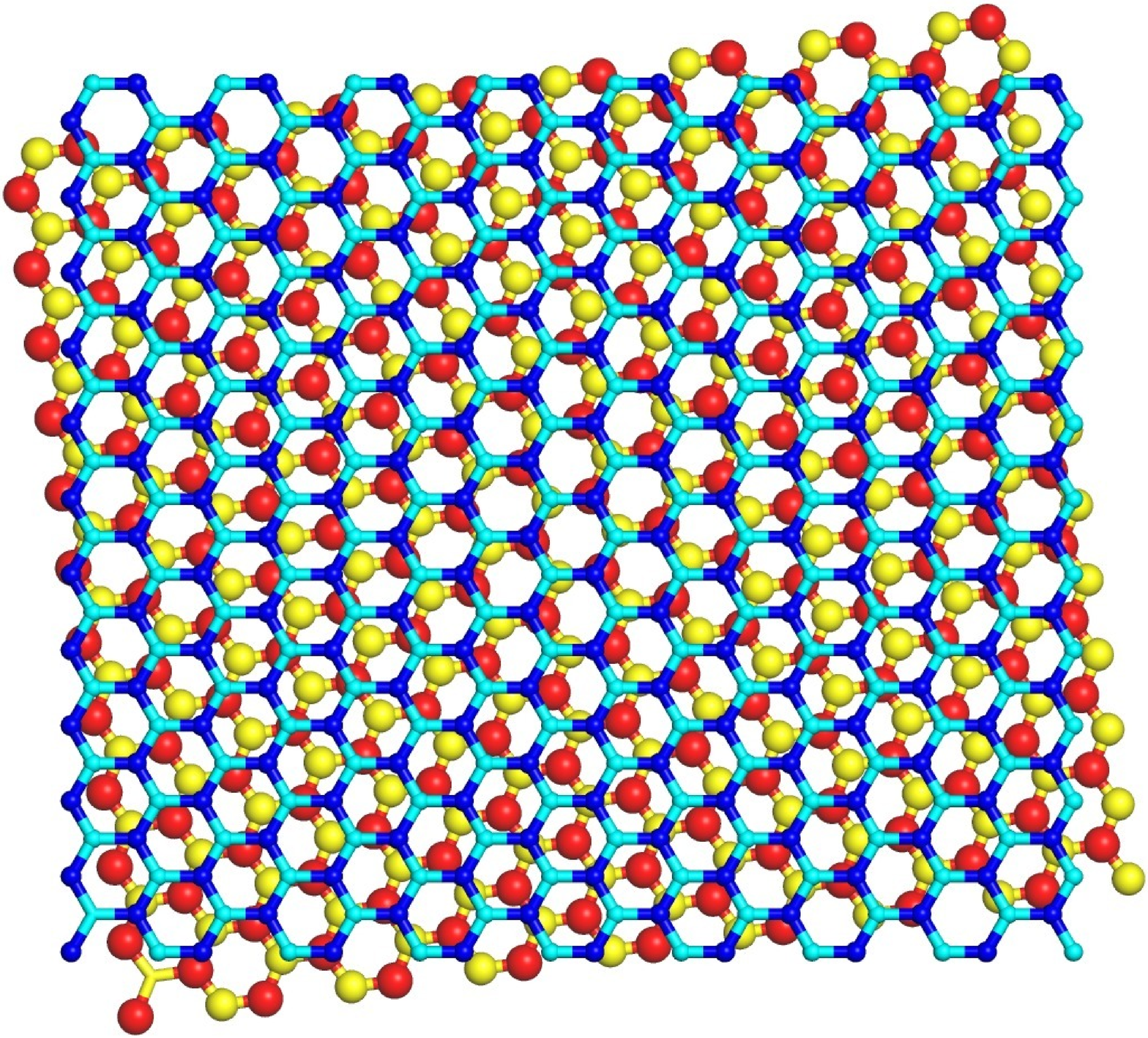}
\caption{(color online) \Moire pattern created by two graphene lattices with a relative twist. Top layer A/B sublattice atoms are shown as small blue/cyan (dark/light) spheres and connectors; bottom layer A/B atoms are shown as large red/yellow (dark/light) spheres. A region of AA alignment lies at the center, where each top-layer atom has a neighbor in the bottom layer. The AA region is surrounded by three AB- and three BA-aligned regions where atoms on only one top-layer sublattice have direct neighbors in the bottom layer. As a consequence, the sublattice-symmetry is broken locally. } \label{fig1} 
\end{figure}

In this Section  we recall the model of Ref.\ \onlinecite{kindermann:prb11}, which underlies also the present article. We analyze the electron dynamics in a graphene layer ``$0$'' when coupled to a second layer ``$1$,'' twisted by a relative angle $\theta$ ($\theta=0^\circ$ for aligned honeycomb lattices, \emph{cf.}\ Fig.~\ref{fig1}), neglecting electron-electron interactions. The corresponding dynamics in multilayers at perturbatively weak interlayer coupling, such as in the experiment \cite{miller:nap10}, are obtained by summation over all layers coupled to the top layer $0$. %, where this allows us to model the interaction between any pair of layers. 
 Twisted graphene bilayers have been described before \cite{lopes:prl07,shallcross:prl08,shallcross:prb10,laissardiere:nal10,mele:prb10} by a tight-binding model with a  local interlayer coupling Hamiltonian  that has parameters fitted to experiment \cite{dresselhaus:aip02},
   \beq
 {H}_{\rm int}= \int {d\boldsymbol{r} \, \Psi^{(0)\dag}(\boldsymbol{r})\Gamma(\boldsymbol{r})\Psi^{(1)}(\boldsymbol{r})+h.c.}
 \eeq
 Here, the spinors $\Psi^{(j)}$  collect the amplitudes for electrons on the two sublattices of layer $j\in\{0,1\}$.  The interlayer coupling  $\Gamma$ has contributions at wavevectors $\boldsymbol{b}^{(0)}-\boldsymbol{b}^{(1)}$, where $\boldsymbol{b}^{(j)}$ are reciprocal  vectors of the graphene lattice in layer $j$ \cite{mele:prb10}.  The Fourier components   of $\Gamma$ quickly decay with increasing wavevector \cite{mele:prb10,shallcross:prl08,shallcross:prb10}.  %A monotonically decreasing function $h(\delta \boldsymbol{r})$ then describes coupling between sites in different layers that decays with increasing distance $ |\delta \boldsymbol{r}|$ between those sites. 
  We therefore neglect all but  the zero wavevector component, setting $\Gamma ({\bf r})=\gamma$. In the ``first star approximation'' of the wavefunctions employed below, the distinction between commensurate and incommensurate interlayer rotations then disappears. %We therefore do not make that distinction.
 This approximation is valid for energies $\varepsilon\gg {\cal V}$, where ${\cal V}$ is set by the Fourier components of $\Gamma$ that directly connect K-points of the two layers \cite{mele:prb10}. We take the limit $0<\theta \ll 1$, when    ${\cal V}\ll\ \gamma$ (in the experiment \cite{miller:nap10} $\theta\approx 0.25^\circ$ and according to the estimate ${\cal V}\simeq \theta^2\gamma$ of Ref.\ \onlinecite{mele:prb10}  this approximation is justified at all accessed energies).

In our limit $0<\theta\ll 1$ a long-wavelength description is appropriate, where the isolated layers $j$ are described by Dirac model Hamiltonians (we set $\hbar =1$)
\beq \label{Dirac} 
 H^{(j)}= \vf \int {d\boldsymbol{r}\sum_\nu \psi_\nu^{(j)\dag}(\boldsymbol{r})\left[
\boldsymbol{\sigma_\nu}\cdot \left(-i
 \boldsymbol{\nabla}+e{\bf A}({\bf r})\right)\right]\psi_\nu^{(j)} (\boldsymbol{r})}.
 \eeq
 Here, $\boldsymbol{\sigma_\nu}=(\nu\sx,\sy)$ is a vector of Pauli matrices, $\nu=\pm $ is the valley spin, $-e$ the electron charge, and $\vf$ the electron velocity
in graphene. We have included an external vector potential ${\bf A}$ to describe a perpendicular magnetic field $B$.
Eq.\ (\ref{Dirac}) acts on the long-wavelength spinors $\psi_{\mu\nu}^{(j)}$ defined by $\Psi^{(j)}_{\mu}(\boldsymbol{r})=\sum_\nu u_{\mu\nu}^{(j)}(\boldsymbol{r})\psi_{\mu\nu}^{(j)}(\boldsymbol{r})$. We write the Bloch functions $u_{\mu\nu}^{(j)}(\boldsymbol{r})=\{\sum_p \exp[i \boldsymbol{K}^{(j)}_{p\nu}\cdot(\boldsymbol{r}-\boldsymbol{\tau}_{\mu}^{(j)})]\}/\sqrt{3}$ in the``first star approximation'' appropriate for the interlayer coupling problem \cite{mele:prb10}. Here, $p$ sums over the three equivalent Brillouin zone corners $\boldsymbol{K}^{(j)}_{p\nu}$ that form the Dirac point of valley $\nu$ \cite{mele:prb10} and   $\boldsymbol{\tau}_{\mu}^{(j)}$  gives the position of an atom on sublattice $\mu \in \{{\rm A,B}\}$ within the unit cell  of layer $j$.  In the long-wavelength theory (which neglects inter-valley processes) the interlayer coupling reads
 \beq
 H_{\rm int} = \int {d\boldsymbol{r} \sum_{\nu} \psi^{(0)\dag}_\nu(\boldsymbol{r})t_\nu(\boldsymbol{r})\psi_\nu^{(1)}(\boldsymbol{r})}+h.c.,
 \eeq
with a matrix $t$ whose long-wavelength components have wavevectors $ \delta\!\boldsymbol{K}_{p\nu} =(R_\theta-1)\boldsymbol{K}^{(0)}_{p\nu}$. Here, $R_\theta$ is a rotation around the $z$-axis by angle $\theta$. Retaining only those long-wavelength parts of $t$ we find
\beq \label{t}
t^{\mu\mu'}_{\nu}\!\!(\boldsymbol{r})\! =\!\frac{\gamma}{3 }\sum_{p}\!e^{i \delta\!\boldsymbol{K}_{p\nu}\cdot\boldsymbol{r}+i\boldsymbol{K}_{p\nu}\cdot\left(\boldsymbol{\tau}_{\mu}^{(0)}-\boldsymbol{\tau}_{\mu'}^{(1)}\right)},
\eeq
where terms of order $\theta$ are neglected, while terms of order $\theta K r$ are kept as they may grow large.

\section{Effective Theory} \label{loc}

\begin{figure}[h]
\subfigure[][~$V^{\rm eff}$]{\includegraphics[width=0.4\linewidth]{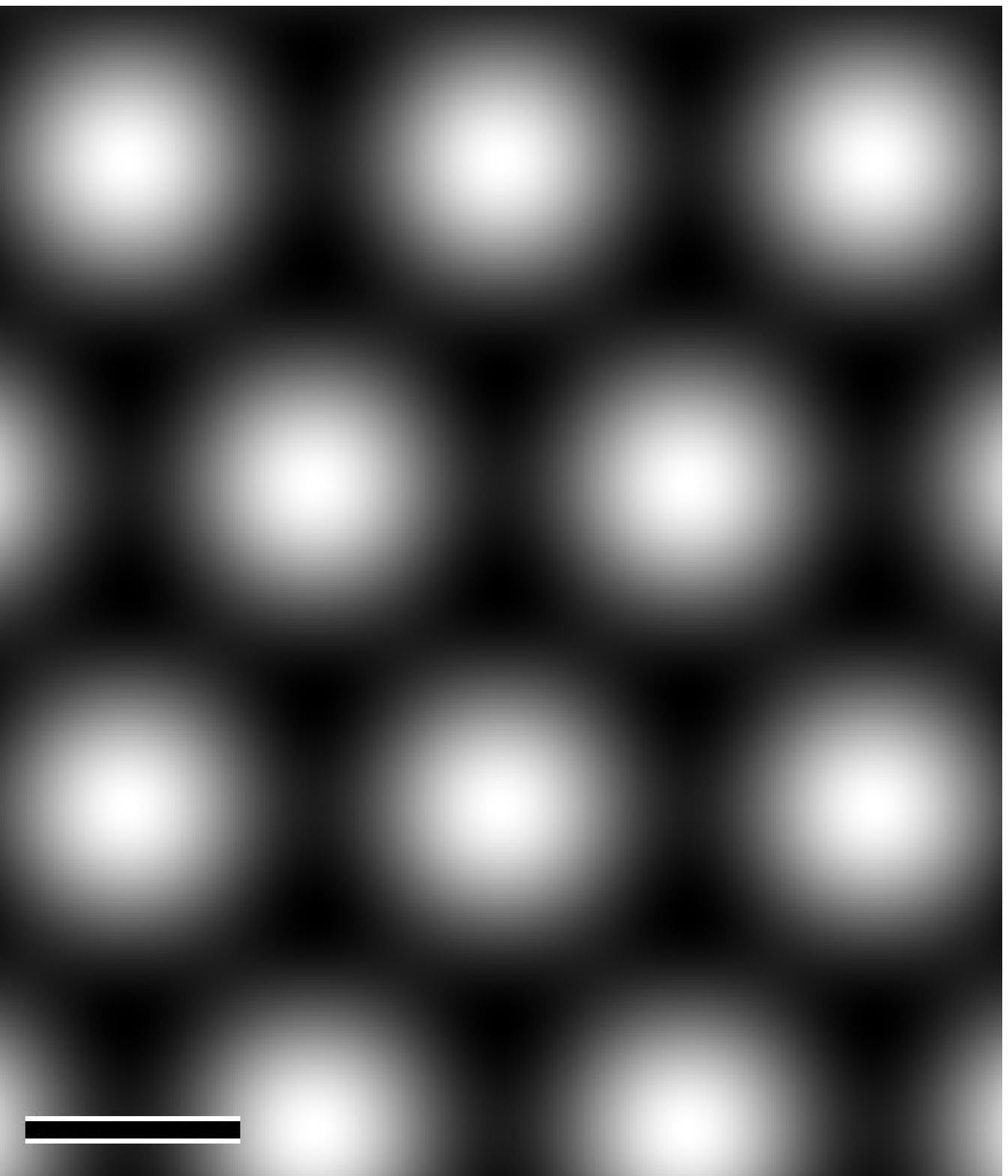}}\hspace{0.08\linewidth}
\subfigure[][~$m^{\rm eff}$]{\includegraphics[width=0.4\linewidth]{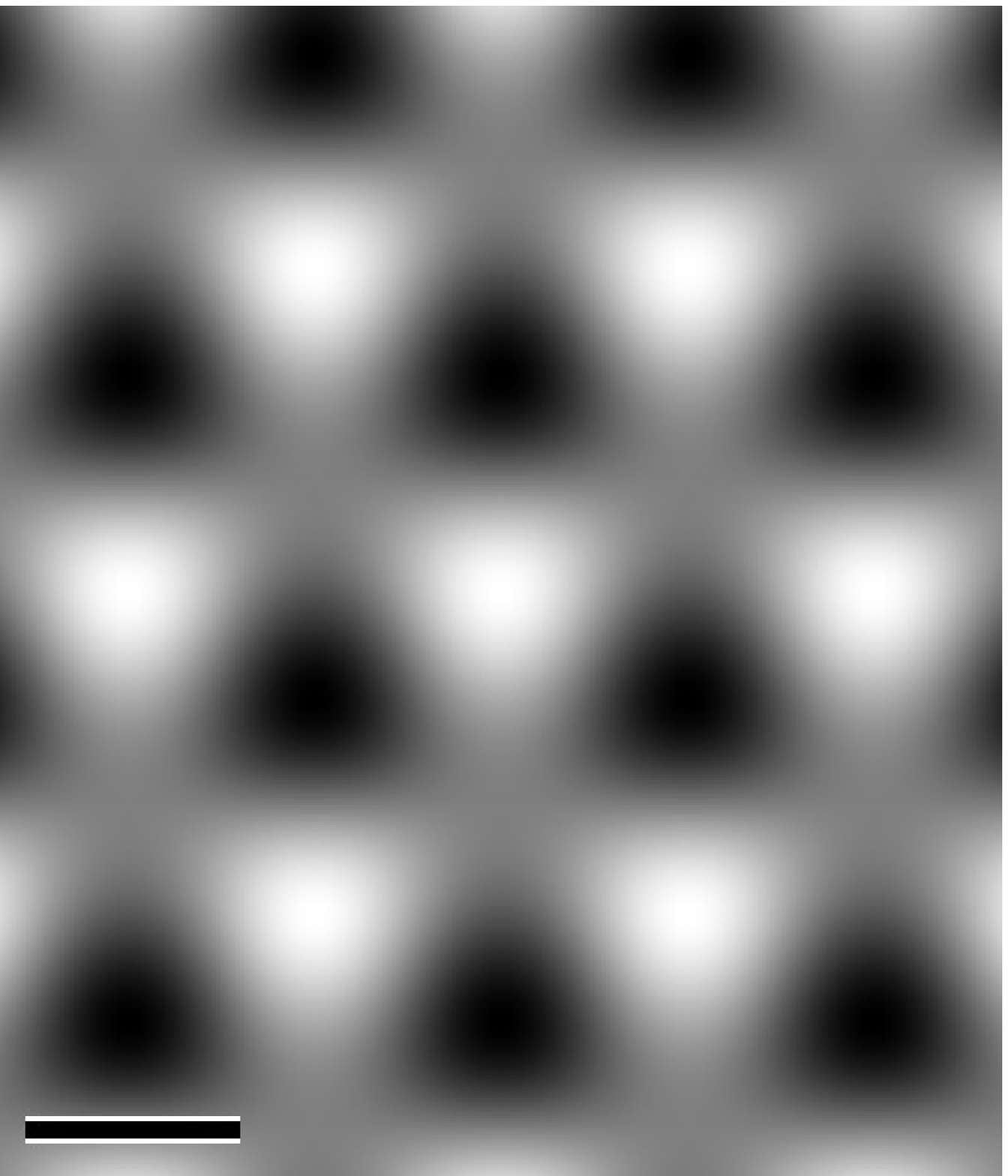}}
\subfigure[][~$A_x^{\rm eff}$]{\includegraphics[width=0.4\linewidth]{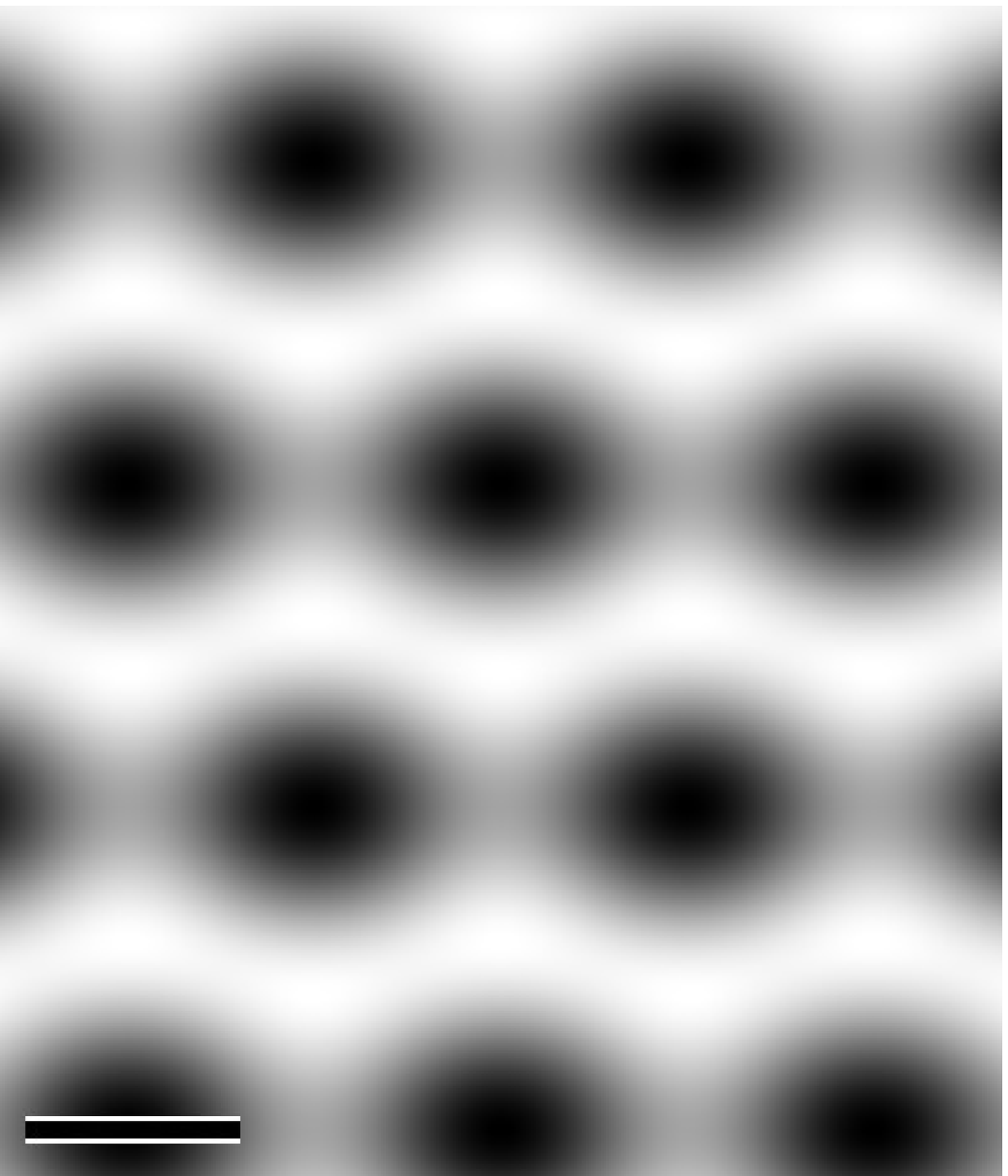}}\hspace{0.08\linewidth}
\subfigure[][~$A_y^{\rm eff}$]{\includegraphics[width=0.4\linewidth]{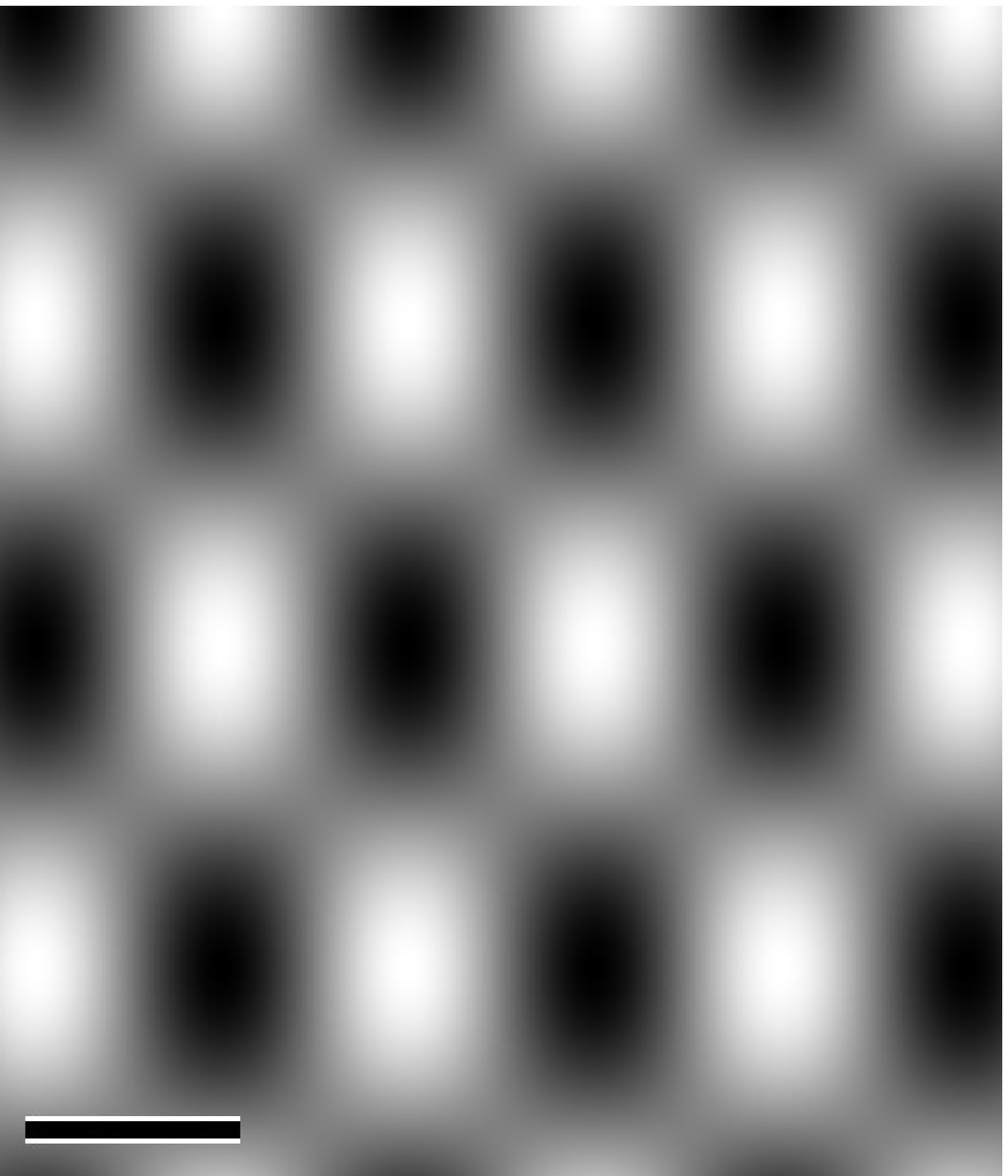}}
\caption{ (a) Effective potential $V^{\rm eff}$, (b) effective mass $m^{\rm eff}$, (c) $A_x^{\rm eff}$, and (d) $A_y^{\rm eff}$ of Eq.\ (\ref{para}) as functions of ${\bf r}\theta/a$ in grey-scale. Scale bars span a unity increment in ${\bf r}\theta/a$, where $a$ is the C-C bond length (0.142 nm). Note the expected sixfold and threefold symmetries of $V^{\rm eff}$ and $m^{\rm eff}$, respectively. ${\bf A}^{\rm eff}$ transforms as a vector under rotations. } \label{fig2} 
\end{figure}

We next integrate out layer $j=1$ in order to arrive at an effective Hamiltonian $H_0^{\rm eff}(\omega)=H_0+\delta\!H_0^{\rm eff}(\omega)$ for the top layer $j=0$, with
\beq \label{Heff}
\delta\!H_0^{\rm eff}(\omega)= H_{\rm int}(\omega+V-H_1)^{-1}H_{\rm int}.
\eeq
We include an interlayer bias $V$ that accounts for different doping levels of the two layers \footnote{In the generalization to multilayers the interaction between layers with $j>0$ needs to be added to the diagonal part $\omega-\sum_j H_j$. }.
In general, $H_0^{\rm eff}$ is nonlocal in space and it depends on the energy $\omega$. In the limit of a large interlayer bias, however, $|V|\gg \omega, \gamma, \theta v/a $, the sum $\omega+V-H_1$ becomes momentum- and energy-independent to a good approximation. The spatial nonlocality and the energy-dependence of $H_0^{\rm eff}$  then may be neglected and $H_0^{\rm eff}$ becomes a conventional Dirac Hamiltonian (\ref{Dirac}) with a matrix potential 
\beq \label{local}
\delta\!H_{0}^{\rm eff}= \int {d\boldsymbol{r}\sum_\nu \psi_\nu^{(0)\dag} (\boldsymbol{r})\frac{t_\nu({\bf r})t_\nu^\dag({\bf r})}{V} \psi_\nu^{(0)} (\boldsymbol{r})},
\eeq
which we parametrize as
\beq \label{para}
\frac{t_\nu({\bf r})t_\nu^\dag({\bf r})}{V}= V^{\rm eff}({\bf r})+\nu \vf e\boldsymbol{\sigma}_\nu \cdot \boldsymbol{A}^{\rm eff}({\bf r})+m^{\rm eff}({\bf r}) \vf^2 \sz.
\eeq
%In our approximation ${\bf b}=0$ in Eq.\ (\ref{t}) we find from Eq.\ (\ref{local})
%\beq
%\delta H_{0\nu; \mu\mu'}^{\rm eff}({\bf r})=-\frac{|\gamma|^2}{9V \Omega^4}\sum_{p, q} e^{i(\delta{\bf K}_p-\delta{\bf K}_q)\cdot {\bf r}+i{\bf K}_p
%\eeq
The interlayer coupling in this limit generates effective scalar and vector potentials $V^{\rm eff}$ and ${\bf A}^{\rm eff}$, respectively, and a mass term $\propto \sz m^{\rm eff} \vf^2$ that implies an effective staggered potential $V^{\rm eff}_{AB}=m^{\rm eff}\vf^2$ in locally Bernal stacked regions.
It follows from Eq.\ (\ref{t}) that $\delta\!H_{0\nu}^{\rm eff}$ oscillates in space with wavevectors ${\bf G}=(R_\theta-1)\boldsymbol{b}$, where ${\bf b}$ is in the ``first star'' of reciprocal lattice vectors of graphene. 
We plot $\delta\!H_{0\nu}^{\rm eff}$ in the parameterization of Eq.\ (\ref{para}) in Fig.\ \ref{fig2}.  The effective Hamiltonian (\ref{local}) promises rich physics. In particular the effective mass term is expected to have profound implications such as topologically confined states \cite{martin:prl08,semenoff:prl08}. In Ref.\ \onlinecite{kindermann:prb11} we have shown that the above effective theory qualitatively and quantitatively accounts for many features of the experiment \cite{miller:nap10}, which was done in a strong magnetic field. In the remainder of this article we explore some of the consequences of the effective potentials (\ref{para}) in zero magnetic field.

\section{Perturbative Results} \label{pert}

\begin{figure}[h]
\includegraphics[width=\linewidth]{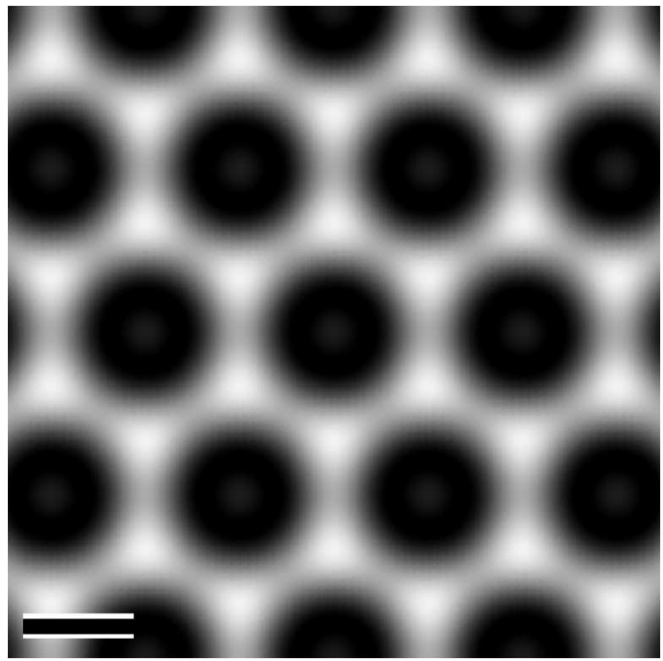}
\caption{  Perturbative correction to the low-energy density of states ($\delta\rho_0/\rho_0$) due to $\delta H_0^{\rm eff}$.  The scale bar corresponds to one unit in ${\bf r}\theta/a$, where $\theta$ is the rotation angle between layers and $a$ is the C-C bond length (0.142 nm).  For a rotation angle of $3^{\circ}$, interlayer coupling of $\unit[\gamma = 300]{meV}$, and an interlayer bias of $\unit[V = 400]{meV}$ (stretching the limits of our locality assumption) the image corresponds to a $\unit[16]{nm}\times\unit[16]{nm}$ area with $\delta\rho_0/\rho_0$ ranging from -0.125   (black) to 0.125   (white). }   \label{fig3} 
\end{figure}

We first explore the perturbative limit of weak interlayer coupling $\gamma^2 \ll V \vf \delta K$ with correspondingly weak effective potentials, Eq.\ (\ref{para}). To this end we do perturbation theory in  $\delta\!H_{0}^{\rm eff}$.    We first obtain the perturbative corrections  to  the low-energy density of states $ \rho_0({\bf r})=\lim_{\varepsilon\to 0} \rho({\bf r},\varepsilon)/\varepsilon$, as probed in  STM measurements.   The lowest order correction to  $ \rho_0({\bf r})=\lim_{\varepsilon\to 0} \rho({\bf r},\varepsilon)/\varepsilon$  vanishes. The leading spatially varying contribution appears at second order in  $\delta\!H_{0}^{\rm eff}$:
\begin{widetext}
\bea\label{drho}
\delta \rho_0({\bf r})&= &\frac{1}{2\pi v^2} \int \frac{d\varphi_k}{2\pi}\sum_{{\bf G},{\bf G}'\neq 0, {\bf G}\neq{\bf G}',s',s''} \lim_{|{\bf k}|\to0} \left[\frac{\langle s,{\bf k},\nu|\delta H_{0\nu}^{\rm eff}| s'', {\bf G}',\nu\rangle \langle s'',{\bf G}',\nu|{\bf r}\rangle\langle {\bf r}| s', {\bf G},\nu\rangle \langle s',{\bf G},\nu|\delta H_{0\nu}^{\rm eff}| s ,{\bf k},\nu\rangle}{\vf^2 |{\bf G}||{\bf G}'|}\right. \nonumber \\
\mbox{} && \left.+ 2 {\rm Re}\frac{ \langle s,{\bf k},\nu|{\bf r}\rangle\langle {\bf r}|s'',{\bf G-G}',\nu\rangle\langle s'',{\bf G-G}',\nu|\delta H_{0\nu}^{\rm eff}| s', {\bf G},\nu\rangle\langle s', {\bf G},\nu|\delta H_{0\nu}^{\rm eff}| s,{\bf k},\nu\rangle}{\vf^2 |{\bf G}| |{\bf G-G}'|}\right].
\eea
Here,   $|s,{\bf k},\nu\rangle$ is an eigenstate of Eq.\ (\ref{Dirac}) at ${\bf A}=0$ with wavevector ${\bf k}$   and energy $s\vf |{\bf k}|$ in valley $\nu$,
\beq
\langle {\bf r}|s,{\bf k},\nu\rangle=\frac{1}{\sqrt{2}} e^{i{\bf k}\cdot{\bf r}}\left(\begin{array}{c} 1 \\ i s\nu e^{i\nu\varphi_k} \end{array} \right),
\eeq
where $\varphi_k=\arctan(k_y/k_x)$. The sums over wavevectors ${\bf G}$, ${\bf G}'$ in Eq.\ (\ref{drho}) runs over all wavevectors contributing to  $\delta H_{0}^{\rm eff}$. Eq.\ (\ref{drho}) evaluated for the effective Hamiltonian (\ref{local}) results in
\beq
\delta \rho_0({\bf r})= \!\!\!\!\!\!\sum_{{\bf G},{\bf G}'\neq 0,{\bf G}\neq{\bf G}'} \!\!\!\!\!\!\!\!\! e^{i({\bf G}-{\bf G}')\cdot {\bf r}}\left[ V^{{\rm eff}*}({\bf G})V^{\rm eff}({\bf G}')+m^{{\rm eff}*}({\bf G})m^{\rm eff}({\bf G}')+{\bf A}^{{\rm eff}*}({\bf G})\cdot {\bf A}^{\rm eff}({\bf G}')\right]\frac{1+(|{\bf G}|+|{\bf G}'|)/|{\bf G}-{\bf G}'|}{2\pi \vf^4 |{\bf G}||{\bf G}'|}.
\eeq
\end{widetext}
 We plot the resulting relative correction to the density of states $\delta \rho_0/\rho_0 = 2\pi \vf^2 \delta \rho_0$   in Fig.\ \ref{fig3}. The result compares well with  the typical \moire patterns observed in STM topography. This suggests that density of states corrections due to the effective potentials Eq.\ (\ref{para}) may  be one of the mechanisms that generate these patterns, besides simple geometric height variations of the top graphene layer (which would be the most straight forward interpretation of topographic STM maps).

We next evaluate the perturbative correction to the electron velocity  at the Dirac point in direction of the momentum $v_{\hat{\bf k}}=\lim_{k\to0}{\bf v}({\bf k})\cdot \hat{\bf k}$, where  $\hat{\bf k}={\bf k}/|{\bf k}|$. One has \cite{park:nap08}
\beq \label{delv}
\delta v_{\hat{\bf k}}=\lim_{|{\bf k}|\to0} \frac{d}{d|{\bf k}|} \sum_{s',{\bf G}}\frac{|\langle s, {\bf k},\nu| \delta H_{0\nu}^{\rm eff}|s',{\bf k-G},\nu \rangle|^2}{\vf(|{\bf k}|-s'|{\bf k-G}|)},
\eeq
 which, for  our $\delta H_{0\nu}^{\rm eff}$  evaluates to
\bea \label{dv}
\delta v_{\hat{\bf k}}&=&-2\vf \sum_{\bf G\neq 0}\Bigl\{ \frac{|V^{\rm eff}({\bf G})|^2[|{\bf G}|^2-(\hat{\bf k}\cdot {\bf G})^2]}{\vf^2 |{\bf G}|^4} \nonumber \\
&&\;\;\;\;\;\;\;\;\;\;\;\;\;\;\;  + \frac{ |m^{\rm eff}({\bf G})|^2(\hat{\bf k}\cdot {\bf G})^2}{\vf^2 |{\bf G}|^4}\Bigr\}.
%\delta v_{\hat{\bf k}}=-2\sum_{\bf G\neq 0}\frac{|V^{\rm eff}({\bf G})|^2\sin^2\theta_{{\bf k,G}}+|m^{\rm eff}({\bf G})|^2\cos^2\theta_{{\bf k,G}}}{\vf |{\bf G}|^2}.
\eea
Here, we have used that ${\bf G}\cdot {\bf A}({\bf G})=0$, which holds in our approximations. % and $ \theta_{{\bf k,G}}$ is the angle enclosed by the vectors ${\bf k}$ and ${\bf G}$.   
 We note that the effective mass suppresses the electron velocity as was found for an oscillating scalar potential $V({\bf r})$ in Ref.\ \onlinecite{park:nap08}. However, differently from a scalar potential, that velocity suppression is in the case of a mass not perpendicular to the direction ${\bf G}$ along which $m^{\rm eff}$ varies, but along that direction.  For the effective Hamiltonian (\ref{local})  [with ${\bf b}=0$ in Eq.\ (\ref{t})]  Eq.\ (\ref{dv}) evaluates to 
 \beq
\delta v_{\hat{\bf k}}=- \frac{|\gamma|^4a^2}{24\pi^2\theta^2V^2},
\eeq
which is  isotropic in space.   We anticipate anisotropic contributions to $\delta v_{\hat{\bf k}}$ at higher orders of perturbation theory. 
The interlayer coupling reduces the velocity, in agreement with earlier calculations \cite{lopes:prl07,laissardiere:nal10} for twisted bilayers at $V=0$. We conclude that in the perturbative regime of weak interlayer coupling the predictions of our theory are consistent with earlier experimental and theoretical work. They moreover have a straightforward interpretation in terms of previous results for  electrons in a superlattice potential \cite{park:nap08}.

\section{Nonperturbative results} \label{nonpert}

 We now turn to the more challenging, but also more interesting nonperturbative limit. For strongly coupled, twisted graphene bilayers a number of  intriguing results have been obtained in electronic structure calculations \cite{laissardiere:nal10,bistritzer:pna11}. For instance, a localization of the electronic wavefunctions in locally AA-stacked regions of the sample and a severe suppression of the electron velocity at small twist angles have been observed  \cite{laissardiere:nal10}. Here, we  show that the mentioned phenomena find an intuitive interpretation in terms of the oscillating mass in a   toy model  of our effective theory. In this toy model we assume  translational invariance in one space direction. Our calculation extends earlier theory of Dirac electrons with a (scalar) superlattice potential  \cite{park:prl09,brey:prl09,arovas:10} to the case of a periodic mass term.

\begin{figure}[h]
\includegraphics[width=\linewidth]{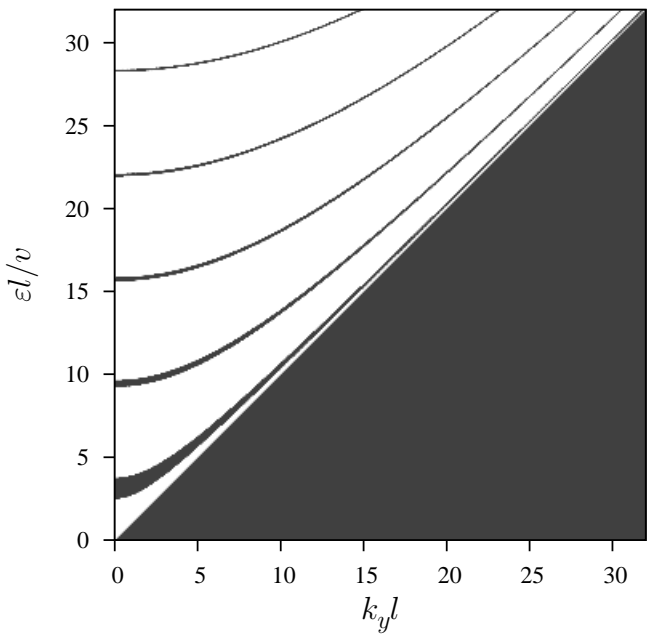}
\caption{ Bandstructure of a Dirac model subject to  a mass term Eq.\ (\ref{1D}) oscillating on a length scale $l$. In the plot, to every  momentum $k_y $, the  energies $\varepsilon$ where electronic states exist are marked white.  For the plot we chose $m=v/l$ }   \label{minibands} 
\end{figure}  

Our toy model   has a mass term, which oscillates in only one space-direction:
 \beq \label{1D}
m^{\rm eff}({\bf r})= m \,{\rm sgn}\,[\cos (x/l)].
\eeq
 Although not directly applicable to the problem of  the interlayer coupling in multilayer graphene%, since the effective Hamiltonian (\ref{local}) varies in both space directions
, such a model is expected to capture some of the same physics.  We  add a   term $m^{\rm eff}\vf^2 \sigma_z$
to Eq.\ (\ref{Dirac}) at ${\bf A}=0$ and find the electronic spectrum along the lines of Ref.\ \onlinecite{arovas:10}. The resulting band structure as a function of the momentum $k_y$ in the direction with translational invariance is plotted in Fig.\ \ref{minibands}. In presence of the periodic mass the electronic spectrum breaks up into  minibands, as for periodic scalar potentials \cite{park:prl09,brey:prl09}. Differently from the scalar case, however, we   find that a periodic mass does not generate new Dirac points, even in the nonperturbative limit $\vf ml\gg 1$. Instead, the bandwidth of the lowest energy band becomes exponentially suppressed  in   $\vf ml$ with correspondingly suppressed  electron velocity: the velocity  in $x$-direction at zero energy is exponentially small in  $\vf ml$,
\beq \label{vx}
v_x|_{\varepsilon=0}=\vf \frac{\vf ml}{\sqrt{2\cosh \vf ml-2}}.
\eeq
The wavefunctions at large $\vf ml$ have dominant weight  around the locations $x=2\pi n$ (with integer $n$) where $m^{\rm eff}$ changes sign. Those are the locations, where for an isolated kink of $m^{\rm eff}$, i.e. a point at which $m^{\rm eff}$ changes sign, topologically protected zero energy states are expected \cite{goldstone:prl81,semenoff:prl08,martin:prl08,yao:prl09}. The lowest energy band observed in Fig.\ \ref{minibands} may be thought of emerging from hybridization of those zero energy states. The larger $\vf ml$ the less overlap occurs between the states localized at adjacent kinks of  $m^{\rm eff}$. This qualitatively explains the exponentially small bandwidth of the lowest energy band in Fig.\ \ref{minibands}. 

The above findings closely resemble the above-mentioned observations of earlier studies of twisted graphene bilayers \cite{laissardiere:nal10}. Also there the wavefunctions were reported to be localized in the AA-stacked regions of the \moire pattern when $\theta\ll1$, which implies large $l$, with $\vf ml\gg 1$. Those are indeed the regions, where $m^{\rm eff}$ in $H^{\rm eff}$, Eq.\ (\ref{para}), changes sign. They thus directly correspond to the regions where the wavefunctions in our toy model Eq.\ (\ref{1D}) are concentrated. Moreover, in Ref.\ \onlinecite{laissardiere:nal10} the electron velocity was found to be strongly suppressed at small $\theta$, corresponding to large $\vf ml$. Also this is in qualitative agreement with the prediction of an exponential suppression of $v_x$ by our model [Eq.\ (\ref{vx})].  

Although our theory is not directly applicable to the calculation of Ref.\ \onlinecite{laissardiere:nal10}, because that calculation was done at $V=0$ and the \moire pattern was naturally two-dimensional, it indeed appears to capture some of its essential physics. The above calculation thus suggests an intuitive interpretation of some of the prominent nonperturbative effects in twisted graphene bilayers in terms of zero energy states that are induced by the topology of the mass term in our   model.

\section{Conclusions}\label{conclusion}

In this article we have discussed some of the implications of the effective theory of rotationally-faulted multilayer graphene that was  put forward in Ref.\ \onlinecite{kindermann:prb11}. We have focused on  its local limit of a large interlayer bias, when this effective theory   takes the form of a conventional Dirac model with space-dependent potentials and mass. While we discussed the implications of that theory for graphene multilayers in a magnetic field in Ref.\  \onlinecite{kindermann:prb11},  here we have explored its consequences in zero magnetic field. In the perturbative limit of weak interlayer coupling we found corrections to the density of states that are consistent with the typical \moire patterns observed in topographic STM measurements. This suggests that these patterns  may not only arise because of height fluctuations, but may at least partially be due to density of states variations. We moreover have found  a velocity correction consistent with earlier calculations in different limits. 

To access the most interesting nonperturbative regime  of strong interlayer coupling we have analyzed a   toy model that   captures most of the essential ingredients of our effective theory. We have demonstrated that that model predicts almost localized electronic states and an exponential velocity suppression. These predictions give an intuitive interpretation to   prior electronic structure calculations for twisted graphene bilayers  in terms of topologically protected zero energy states localized at kinks of an oscillatory Dirac mass term. Partially answering the question we raised at the outset,   the presented calculations lead us to the following conclusion:  while our real-space theory of the interlayer coupling certainly is an advantageous description of twisted graphene bilayers with  a large interlayer bias, it   qualitatively captures much of their  essential physics  even without such bias, when it does not strictly apply. The theory of Ref.\  \onlinecite{kindermann:prb11} indeed appears to be an advantageous approach to the physics of twisted graphene bilayers, also in zero magnetic field.

\acknowledgements

We gratefully acknowledge discussions with W.\ de Heer, E.\ J.\ Mele, D.\ L.\ Miller. This work was funded in part by the NSF (DMR-1106131 and DMR-0820382) and by the Semiconductor Research Corporation Nanoelectronics Research Initiative (NRI-INDEX).

 \end{document}